\documentclass[twocolumn]{aastex631}

\begin{document}

\title{A Sizable Discrepancy in Ground-Based JAGB Distances to Nearby Galaxies}

\author[0000-0002-5259-2314]{Gagandeep S.~Anand}
\affiliation{Space Telescope Science Institute, 3700 San Martin Drive, Baltimore, MD 21218, USA}

\begin{abstract}

Recently, \cite{2024arXiv240806153F} report agreement of distances derived from the Tip of the Red Giant Branch (TRGB) and the J-Region Asymptotic Giant Branch (JAGB) at the 1$\%$ level for both nearby galaxies with ground-based imaging (0.5$-$4~Mpc) as well distant galaxies with JWST imaging (7$-$23~Mpc). Here we compare the same ground-based JAGB distances to uniformly reduced space-based optical TRGB distances from the Hubble Space Telescope (HST). We uncover a significant offset between these two distance scales of $\Delta\mu$ = 0.17 $\pm$ 0.04 (stat) $\pm$ 0.06 (sys)~mag (9$\%$ in distance), with the HST TRGB distances being further. Inspections of the HST color-magnitude diagrams make a compelling case that the issue lies in the underlying JAGB distances. The source of the disagreement may lie with the lower resolution or photometric calibration of the ground-based near-infrared data, a contrast to the general agreement found between JWST JAGB and other space-based, second-rung distance indicators (Cepheids, Miras, TRGB) presented within \cite{2024arXiv240811770R}. High-resolution, near-infrared observations from an ongoing HST program will enable the simultaneous measurement of Cepheid, JAGB, and TRGB distances in four of these nearby galaxies and allow us to investigate whether the discrepancy noted here is due to ground-based observational systematics, or something intrinsic to the JAGB method relevant for this particular sample. A resolution of this discrepancy is required if the JAGB is to be used to determine a highly precise local value of the Hubble constant.

\end{abstract}

\keywords{Carbon stars; Distance indicators; Red giant tip; Stellar distance}

\section{Introduction} \label{sec:intro}

\cite{2024arXiv240806153F} have recently claimed that the tip of the red giant branch (TRGB; \citealt{1993ApJ...417..553L, 1995AJ....109.1645M, Beaton2018, 2021AJ....162...80A}) and J-region asymptotic giant branch (JAGB; \citealt{2001ApJ...548..712W, 2020ApJ...899...66M, 2024ApJ...966...20L}) distance scales are in excellent agreement (1$\%$), both locally (0.5$-$4~Mpc; \citealt{2020ApJ...899...67F, 2024ApJ...967...22L}), and in more distant galaxies observed with JWST (7$-$23~Mpc; \citealt{2024ApJ...961..132L, 2024arXiv240803474L}). Here we compare ground-based JAGB distances from \cite{2024ApJ...967...22L} to space-based optical TRGB measurements from the Extragalactic Distance Database's (EDD) CMDs/TRGB catalog\footnote{Available at \url{edd.ifa.hawaii.edu}.} \citep{2009AJ....138..323T, 2009AJ....138..332J, 2021AJ....162...80A}. A key benefit of this limited comparison is that both sets of photometry and distances are publicly available, and measured with the same observatory within each dataset (the JAGB imaging was taken with Magellan-FourStar, whereas the TRGB imaging was taken with the Hubble Space Telescope).

\begin{figure*}
\epsscale{1.1}
\plotone{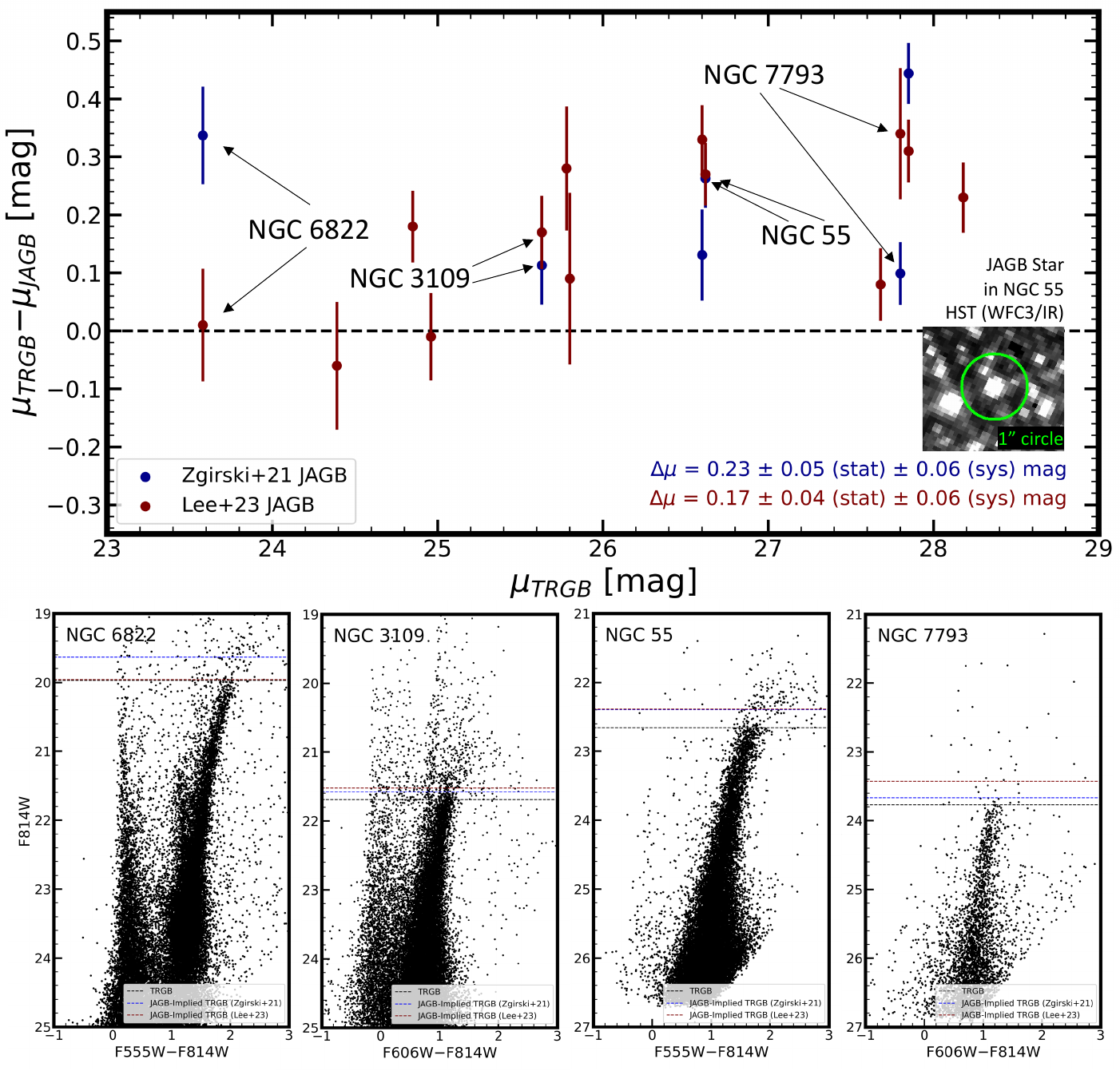}
\caption{\textbf{Top:} Comparison of ground-based JAGB \citep{2024ApJ...967...22L} and HST optical TRGB distance moduli \citep{2009AJ....138..332J, 2021AJ....162...80A} shown in red. We find a significant offset of $\Delta\mu$ = 0.17 $\pm$ 0.04 (stat) $\pm$ 0.06 (sys)~mag between the two samples. The blue points show the same comparison with JAGB distance moduli from \cite{Zgirski_2021ApJ...916...19Z}, but with only six overlapping targets. A postage stamp of a JAGB star in the outer disk of NGC~55 observed with HST WFC3/IR is shown in the lower-right$–$ the green circle has a 1$''$ radius for scale, suggesting that ground-based photometry of the same object is likely to suffer from the effects of crowding. \textbf{Bottom:} Individual HST CMDs for a representative range of underlying galaxies in the sample, sorted by increasing distance. The original TRGB measurements are given in black, whereas the JAGB-implied TRGB values are given in red and blue (the latter only if available). It can be readily seen that some of the JAGB distances are not easily reconciled with the visible termination of the red giant branch in these nearby (0.5$-$4~Mpc) galaxies.}
\label{fig:compare}
\end{figure*}

\section{Comparison of Distance Scales}

The comparison between the two samples is shown in the top-half of Figure \ref{fig:compare}. The plotted uncertainties in the JAGB distances are re-tabulated from Table 3 in \cite{2024ApJ...967...22L} to include contributions from the 2MASS photometric zeropoint and 2MASS-to-FourStar correction, error on the JAGB mode, choice of smoothing scale, and a 15$\%$ uncertainty in the foreground reddening ($A_{J}$). We also include JAGB results for WLM \citep{2021ApJ...907..112L} and M33 \citep{2022ApJ...933..201L}, which are part of the same ground-based observing program, for a total of thirteen galaxies for comparison. The uncertainties in the TRGB distances are re-tabulated from the CMDs/TRGB catalog to include the statistical uncertainty in the tip measurement and the same 15$\%$ uncertainty in foreground reddening ($A_{F814W}$). The mean difference between the two sets of distances is $\Delta\mu$ = 0.17 $\pm$ 0.04 (stat) $\pm$ 0.06 (sys)~mag. The statistical uncertainty is given by $\sigma/\sqrt{N}$, and the systematic uncertainty is given by the quadrature sum of the underlying calibration uncertainty for each method (0.04 mag for each; \citealt{Rizzi2007,2021ApJ...919...16F, 2024ApJ...967...22L}). Separately, the blue points show a comparison between the EDD TRGB distances and the ground-based JAGB distances from \cite{Zgirski_2021ApJ...916...19Z} using data from the InfraRed Survey Facility (IRSF), which show a somewhat increased discrepancy of $\Delta\mu$ = 0.23 $\pm$ 0.05 (stat) $\pm$ 0.06 (sys)~mag, but with only six overlapping galaxies.

The bottom-half of Figure \ref{fig:compare} shows four example HST CMDs used for the TRGB measurements. On each of these CMDs, we plot the measured magnitude of the TRGB (in black), and the implied magnitude of the TRGB given by the JAGB distances (in red and blue). For NGC~6822, we see excellent agreement between the TRGB measurement from \cite{2021AJ....162...80A} and the JAGB value from \cite{2024ApJ...967...22L}. However, for the majority (3/4 cases shown, 9/13 cases total), we see that the JAGB-implied values are  inconsistent with the visible onset of the TRGB in these nearby galaxies. While only four examples are shown here, the remainder of the HST CMDs are available via EDD.

\section{Potential Solutions}

The underlying zeropoint and color-calibration used by the EDD CMDs/TRGB catalog is provided by \cite{2007ApJ...661..815R}, with a baseline value of $M_{F814W}$~=~$-4.05$~mag. The average zeropoint for the galaxies in question after including the color calibration is $M_{F814W}$~=~$-4.04$~mag, in good agreement with prior results from the Carnegie-Chicago Hubble Program who find $M_{F814W}$~=~$-4.05$~$\pm$~$0.04$~mag \citep{2021ApJ...919...16F}. Given the level of agreement, it seems unlikely that TRGB zeropoint is responsible for the discrepancy.

Could the HST photometry or measurements contained within EDD be subject to large, systematic offsets? Fortunately, subsamples of these 13 galaxies have HST TRGB measurements from other groups, with distinct photometry and TRGB measurement pipelines. The ANGST team \citep{2009ApJS..183...67D} measured distances to six of the thirteen galaxies compared here, as was also noted by \cite{2024ApJ...967...22L}. Using the ANGST TRGB values collated by \cite{2024ApJ...967...22L}, the average distance between the JAGB and TRGB distance moduli is $\Delta\mu$~=~0.21~$\pm$~0.04 (stat)~mag, very similar to the offset found in the full comparison using the EDD TRGB sample. Similarly, the GHOSTS team \citep{2011ApJS..195...18R} has measured a TRGB distance to NGC 7793 in three separate fields, and find a mean distance modulus of $\mu$ = 27.87~mag, resulting in an even larger discrepancy between TRGB and JAGB distance scales ($\Delta\mu$~=~0.41~mag) than found with the EDD TRGB measurement ($\Delta\mu$~=~0.34~mag). In short, the HST TRGB distances from the separate groups are all consistently offset in comparison to this sample of JAGB distances to nearby galaxies.

\cite{2024ApJ...967...22L} mention crowding as a potential cause of concern for NGC 1313, the furthest target in their sample (D~=~4.3~Mpc). However, we see large offsets in derived distances beginning at just over 1~Mpc. Fortuitously, an ongoing HST observing program (GO–17520; \citealt{2023hst..prop17520B}) is obtaining high-resolution imaging of four of the galaxies in this sample (NGC~55, NGC~300, NGC~3109, NGC~6822) for the express purpose of performing a 1$\%$ cross calibration of the Cepheid, JAGB, and TRGB distance indicators. A WFC3/IR postage stamp of a JAGB star in the outer disk of NGC~55 from this program is shown in the lower-right hand corner of the top panel of Figure \ref{fig:compare}. A green circle with 1$''$ radius is overplotted, and suggests that the ground-based photometry may suffer from the effects of stellar crowding, even in the outer disk regions utilized by \cite{2024ApJ...967...22L}. 

In addition to potential crowding, we find that there may be an issue with the photometric calibration of the ground-based data. \cite{2024ApJ...967...22L} also present near-infrared TRGB distances to many of the same galaxies with the same set of photometry. For eleven galaxies in common, we find a difference of $\Delta\mu$ = 0.12 $\pm$ 0.04 (stat) $\pm$ 0.07 (sys)~mag between our optical HST TRGB and their ground-based NIR-TRGB distances. This disagreement is in the same direction as the disagreement between the HST TRGB and the ground-based JAGB measurements, albeit at a somewhat reduced level (and with less significance). Given that the TRGB stars used by \cite{2024ApJ...967...22L} lie in significantly less dense regions than their JAGB stars and are thus less susceptible to crowding, the underlying photometric calibration may also be a culprit.

It seems likely that crowding and/or photometric calibration are a significant part of the solution, but it is worth pointing out that the nearby galaxies in the comparison presented here are somewhat lower mass than those used in the Cepheid+Type Ia distance ladder \citep{2022ApJ...934L...7R}. For those galaxies, there is good general agreement between JWST JAGB measurements and those from other second-rung distance indicators \citep{2024ApJ...966...20L, 2024arXiv240803474L, 2024arXiv240811770R}. Notably, JWST imaging in the 7-23~Mpc range is of higher resolution than ground-based imaging in the 2-4~Mpc range. Future investigations into the dependence of the JAGB on metallicities, ages, and full star-formation histories are duly warranted \citep{2023ApJ...956...15L, 2024ApJ...966...20L}. 

\begin{acknowledgments}
G.S.A thanks Abigail Lee, Adam Riess, Rachael Beaton, Louise Breuval, and Siyang Li for useful conversations. 
\end{acknowledgments}

\bibliography{paper}{}
\bibliographystyle{aasjournal}

\end{document}